\documentclass[11pt]{article}

\usepackage[number]{achemso}
\usepackage{graphicx}% Include figure files

\newcommand{\dg}{$\Delta G$}
\newcommand{\ds}{$\Delta S$}

\begin{document}

\title{Peptide Conformational Equilibria Computed via a Single-Stage
Shifting Protocol}
\author{F.\ Marty Ytreberg\footnote{E-mail: fmy1@pitt.edu}
and Daniel M.\ Zuckerman\footnote{E-mail: dmz@ccbb.pitt.edu}\\
Computational Biology, School of Medicine\\
Dept.\ of Environmental and Occupational Health,\\
Graduate School of Public Health,\\
University of Pittsburgh, 200 Lothrop St., Pittsburgh, PA 15261}
\date{\today}
\maketitle

\begin{abstract}
We study the conformational equilibria of two peptides using a novel
statistical mechanics approach designed for calculating free energy
differences between highly dis-similar conformational states.
Our results elucidate the contrasting roles of entropy in
implicitly solvated leucine dipeptide and decaglycine.
The method extends earlier work by Voter, and
overcomes the notorious ``overlap'' problem in free energy
computations by constructing
a mathematically equivalent calculation with high conformational
similarity.
The approach requires only equilibrium simulations of
the two states of interest,
without the need for sampling transition states.
We discuss extensions of the approach to binding
affinity estimation
and explicitly solvated systems, as well as possible
optimizations.
\end{abstract}

\section{Introduction}
Although free energy differences (\dg) are
fundamental to the description of every
molecular process, computer-based estimation
of \dg\ remains among the most difficult and time-consuming tasks
in computational chemistry and biology.
Despite the obstacles, molecular mechanics
\dg\ calculations are presently applied for protein engineering
\cite{degrado,mayo}, drug design \cite{jorgensen-sci,burgers},
toxicology studies \cite{vangunsteren-estrogen},
solubility estimation \cite{grossfield-jacs,vangunsteren-onestep},
and determining binding affinities of ligands to
proteins \cite{kollman-pnas}.
Although many {\it ad hoc} and approximate methods are
in wide use \cite{aqvist-febs}, including docking
protocols \cite{shoichet-nature,shoichet,nussinov,kuntz,scheraga},
there is a widely recognized
need to more rigorously include molecular flexibility
and entropic effects \cite{shoichet-nature,nussinov,kuntz,scheraga,shoichet}.
Additionally, the development of improved, polarizable molecular
mechanics forcefields
\cite{brooks-polar,berne-polar,roux-polar,grossfield-jacs}
suggests rigorous,
ensemble-based \dg\ estimates will play an increasing role 
in elucidating subtle molecular effects---aided by the
decreasing cost of computer resources.

Our concern here is to calculate the conformational
free energy differences \dg\ and entropy differences
\ds\ for peptides
with a method based on statistical mechanics,
which fully accounts for molecular flexibility. Because
these techniques attempt a full accounting of entropic
and enthalpic effects, they are computationally
demanding---and it is the need for greater efficiency which
we address here. Our ideas are framed and tested for molecular
mechanics calculations, but they should also prove applicable
with quantum mechanical computation \cite{karplus-qm}.

Present statistical mechanics methods can be classified into equilibrium
and non-equilibrium approaches.
Equilibrium \dg\ calculations, such as thermodynamic integration
and multi-stage free energy perturbation, have been used
successfully for many years
\cite{karplus-jcp,mccammon,jorgensen,shirts-jcp}.
Equilibrium methods can be very accurate, but have a large
computational cost.
On the other hand, Jarzynski's equality \cite{jarzynski}
has recently has opened up a host
of {\it non}-equilibrium \dg\ methods
\cite{zuckerman-prl,hummer-pull,bustamante-sci,schulten-pmf,schulten-steered}.
Non-equilibrium methods can provide rapid estimates for \dg, but typically
suffer from bias without careful convergence testing
\cite{zuckerman-cpl,ytreberg-seps,ytreberg-extrap,hummer-fep}.
Single-stage approaches \cite{mccammon-ti,vangunsteren-onestep},
may be considered non-equilibrium calculations which are based upon
simulating only one or both end states of interest.

Poor ``overlap''---the lack of conformational similarity between the
molecular states
of interest---is a key cause of computational expense in
statistical mechanics methods.
This phenomenon is illustrated schematically in Figure \ref{fig-tors}.
Unfortunately, in many cases of interest, poor overlap is the rule
rather than the exception.
Such dis-similarity is implicit in the critical problem of
conformational equilibria, for instance.

The most common approach to improve overlap in free energy
calculations is that used in thermodynamic integration, namely,
simulating the system at
multiple hybrid, intermediate stages (e.g., Refs.\
\citenum{grossfield-jacs,beveridge,jorgensen,lu-jcc,karplus-jcp,mccammon,shirts-jcp}).
An alternative strategy was taken by
van Gunsteren and collaborators \cite{vangunsteren-ref2} and
McCammon and collaborators \cite{mccammon-ti}
who built non-physical intermediate reference states that
increased overlap between the two end states.
A third noteworthy solution to the overlap problem is the
computation of absolute free
energies for each of the end states, avoiding any need for
configurational overlap
\cite{meirovitch-deca,gilson-jacs,karplus-deca,stoessel}.

Here, we address the overlap problem using a
two-decade-old, elegant idea from solid-state systems \cite{voter}.
In Ref.\ \citenum{voter}, Voter pointed out
that shifting potential energy functions (or coordinates)
by a constant vector in configuration space
can dramatically improve the overlap of the states
while maintaining the value of \dg\ {\it exactly};
he applied this idea to tungsten dimers on a crystal surface.
A similar approach was adopted for lattice systems by Bruce
and collaborators \cite{bruce}.
In their ``metric scaling'' approach, Reinhardt
and collaborators also built on Voter's idea for
slow-growth \dg\ estimates
for crystal lattice changes \cite{reinhardt-scaling}.
Most recently, in a generalization of Voter's approach,
Jarzynski introduced ``targeted free energy perturbation,''
and applied the method to a Lennard-Jones fluid
\cite{jarzynski-targeted}.
However, Ref.\ \citenum{jarzynski-targeted}
also indicates that constructing the mapping is
likely to be as hard as estimating the free energy itself, and

Here, we introduce the first generalization of Voter's shifting
strategy specifically tailored to
molecular calculations.
Because no intermediate stages are required,
the approach has the potential to dramatically reduce computation times.
Overlap is obtained by shifting
{\it internal coordinates} of the molecule.
The shifting strategy is further combined with
Bennett's iterative approach \cite{bennett} to efficiently utilize the
data. These two ideas, (i) shifting internal coordinates
and (ii) the use of Bennett's iterative method, provide the backbone of the
single-stage shifting approach presented here.

In outline, this report first builds the necessary mathematical framework
for the single-stage shifting method in Section \ref{sec-sss}.
Then in Section \ref{sec-results}, the method is tested on
leucine dipeptide (ACE-(leu)$_2$-NME) in GBSA implicit solvent, correctly
calculating the free energy difference \dg\ and
entropy difference \ds\ between the alpha and beta conformations.
The leucine dipeptide system is also used to demonstrate various
shifting approaches, and show the importance of using
Bennett's iterative method. Finally, the single-stage
shifting method is used to
predict \dg\ and \ds\ between the alpha and
extended conformations
of decaglycine (ACE-(gly)$_{10}$-NME) in GBSA solvent. 

While our results already demonstrate notable efficiency (the decaglycine
calculations, for instance, would be extremely costly by
conventional staging methods), we have not yet
pursued a number of fairly clear avenues for
optimization; these are discussed in Section \ref{sec-opt}.
Finally, the extension of the single-stage shifting method to
explicitly solvated systems, and for ``alchemical''
mutations, is explored in Section \ref{sec-ext}.

\section{\label{sec-sss}The single-stage shifting method}
In this section we introduce a single-stage method
which improves the overlap between
the end states by construction. In essence, a mathematically
equivalent calculation, with high overlap, is performed
instead of addressing the original problem.

\subsection{Constructing an equivalent shifted \dg\ calculation}
Consider two end states defined by potential energy functions
$U_0(\vec{x})$ and $U_1(\vec{x})$, where $\vec{x}$
is a set of configurational coordinates. The
two states could be two different conformations, or
the bound and unbound states of a protein.
For problems in conformational equilibria, as studied below,
$U_0$ and $U_1$ can be the same potential function restricted
to different regions of configurational space (i.e., two conformational
states).
The free energy difference \dg\ between these two states
is given by
\begin{eqnarray}
    e^{-\beta \Delta G}=\frac{Z\bigl[U_1(\vec{x})\bigr]}
			     {Z\bigl[U_0(\vec{x})\bigr]}=
	\frac{\int d \vec{x} \; e^{-\beta U_1(\vec{x})}}
	     {\int d \vec{x} \; e^{-\beta U_0(\vec{x})}},
    \label{eq-z1z2}
\end{eqnarray}
where $\beta=1/k_BT$, and $Z$ indicates a partition
function, and integration is to be performed over the indicated
potential function.
Note that for implicit solvation, as is used in this study, the
Gibbs free energy is equivalent to the Helmholtz free energy.
Using the formalism of free energy perturbation \cite{zwanzig},
we can re-write eq (\ref{eq-z1z2}) as
\begin{eqnarray}
    e^{-\beta \Delta G}=
	\frac{ \int d \vec{x} \; \Bigl(
		e^{-\beta \bigl[ U_1(\vec{x})-U_0(\vec{x}) \bigr]} \Bigr)
		\; e^{-\beta U_0(\vec{x})} }
	     {\int d \vec{x} \; e^{-\beta U_0(\vec{x})}}
	= 
    \Biggl< e^{-\beta \bigl[ U_1(\vec{x})-U_0(\vec{x}) \bigr]} \Biggr>_0,
    \label{eq-fep1}
\end{eqnarray}
where the $\langle ... \rangle_0$ indicates an equilibrium
average over $U_0$ configurations.

Voter's shifting strategy \cite{voter} improves the overlap between
states (potential functions) without altering \dg.
Imagine shifting the origin of the configurational coordinates $\vec{x}$
in $U_1$ by a constant vector $\vec{C}$. This corresponds to
a simple change of variables,
$\vec{x} \rightarrow \vec{x}+\vec{C}$, and
$Z\bigl[U_1(\vec{x}+\vec{C})\bigr]=Z\bigl[U_1(\vec{x})\bigl]$
remains unchanged
because integration is performed over all space.
Eq (\ref{eq-fep1}) can now be written as
\begin{eqnarray}
    e^{-\beta \Delta G}=
    \frac{ \int d \vec{x} \; \Bigl(
	e^{-\beta \bigl[ U_1(\vec{x}+\vec{C})-U_0(\vec{x}) \bigr] } \Bigr)
	\; e^{-\beta U_0(\vec{x})} }
         {\int d \vec{x} \; e^{-\beta U_0(\vec{x})}}
    = \Biggl<
	e^{-\beta \bigl[ U_1(\vec{x}+\vec{C})-U_0(\vec{x}) \bigr]}
    \Biggr>_0.
    \label{eq-fep}
\end{eqnarray}
In principle, eq (\ref{eq-fep}) implies that one can
arbitrarily shift the coordinates for the $U_0$ configurations
needed for the average.
(Shifting $U_0$ rather than $U_1$ produces an analogous result.)
In practice, the shifting should be done to
maximize the overlap between configurations in $U_0$ and $U_1$.
Unfortunately, no {\it Cartesian} shift vector can bring two distinct
molecular configurations into overlap. Therefore, we apply
the Voter idea in {\it internal coordinates}---and
we term the approach single-stage shifting.

\begin{figure}
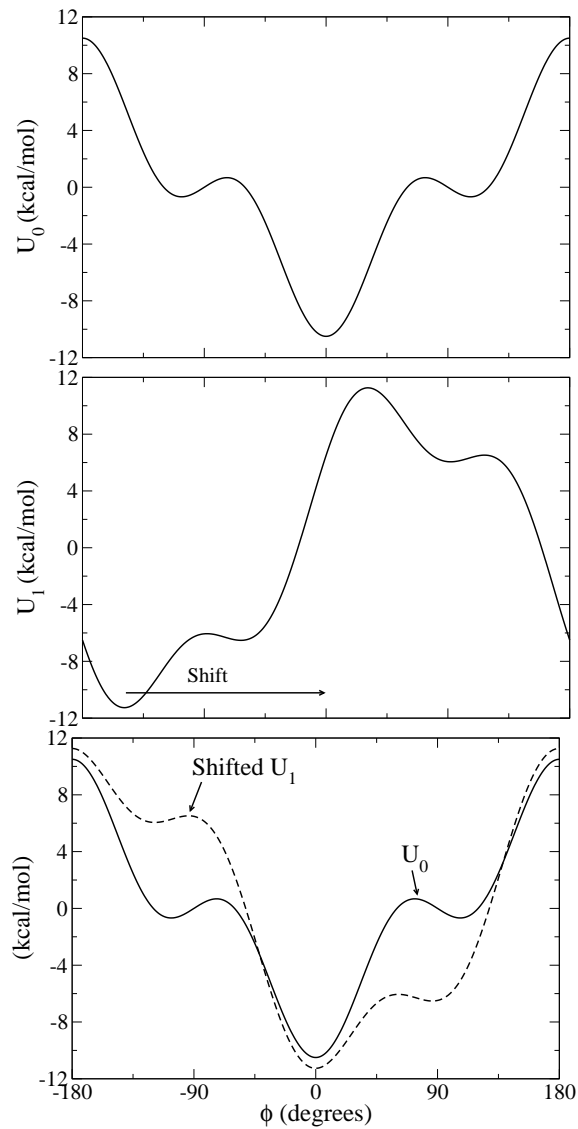

    \begin{center}
	\includegraphics[scale=0.3,clip]{tors-sketch-U0.eps}\\
	\includegraphics[scale=0.3,clip]{tors-sketch-U1.eps}\\
	\includegraphics[scale=0.3,clip]{tors-sketch-shift.eps}
    \end{center}
    \caption{\label{fig-tors}
	Illustration of the shifting idea using two idealized
	torsional potentials (a) $U_0(\phi)$
	and (b) $U_1(\phi)$. There is minimal configurational overlap
	between these two potentials, since simulations will mainly
	sample the deep, dis-similar minima of the potentials.
	(c) With the use of an appropriate shifting constant, it is possible
	to construct overlap between the states without altering
	the numerical value of the conformational free energy difference.
	Note that for peptides, we shift
	internal coordinates rather than potentials
	(see Section \ref{sec-use}), however, these two
	procedures are equivalent.
    }
\end{figure}

To demonstrate the reasoning behind a shift in internal coordinates,
consider the schematic torsional potentials shown in
Figure \ref{fig-tors}a,b.
Simulations for $U_1$ will mainly sample
the large ``trans'' well at $\phi=0$, while simulations for
$U_0$ will mainly sample the large ``gauche'' well at
$\phi\approx-150$. Shifting $U_1$ by a constant corresponding
to the difference between the minima of the
two potentials (i.e., roughly 150 degrees for this example)
produces Figure \ref{fig-tors}c,
where the overlap between $\phi$ configurations is excellent.
Such a shift does {\it not} alter the $U_1$ partition function
and thus \dg\ is unaffected.
Note that, below, we shift internal coordinates rather than
potentials, but these are equivalent procedures.

In practice, the dimensionality of the system will be high and
the shifting constant will not be as obvious as that in Figure \ref{fig-tors}.
Reasonable methods for determining the shifting constant will
be discussed in Section \ref{sec-use}.

\subsection{Utilizing bi-directional data---Bennett's Methods}
In single-stage free energy perturbation (eq (\ref{eq-fep})),
only configurations from $U_0$ are evaluated using $U_1$.
However, if simulations are performed in both states
of interest, one could just as
readily evaluate configurations from $U_1$ using $U_0$.
Bennett showed that the ``bi-directional''
($U_0 \rightarrow U_1$ and $U_1 \rightarrow U_0$) evaluations could
be combined to minimize the uncertainty in \dg
\cite{bennett}.

Bennett introduced both ``iterative'' and ``acceptance-ratio''
methods to utilize bi-directional data \cite{bennett}.
Generalizing the acceptance-ratio formulation
to the shifting approach yields \cite{voter,bennett}
\begin{eqnarray}
    e^{-\beta \Delta G}=
    \frac{
	\Biggl<
	    \min \Bigl(
		1.0,e^{-\beta \bigl( U_1(\vec{x}+\vec{C})-U_0(\vec{x}) \bigr)}
	    \Bigr)
	\Biggr>_0
    }
    {
	\Biggl<
	    \min \Bigl(
		1.0,e^{-\beta \bigl( U_0(\vec{x}-\vec{C})-U_1(\vec{x}) \bigr)}
	    \Bigr)
	\Biggr>_1
    }.
    \label{eq-ar}
\end{eqnarray}
Similarly, generalizing to the iterative method to shifted
coordinates leads to the following relation \cite{bennett}
\begin{eqnarray}
    \Biggl< \biggl(
	1+e^{-\beta \bigl( U_0(\vec{x})-U_1(\vec{x}+\vec{C})+\Delta G \bigr)}
	\biggr)^{-1}
    \Biggr>_0=
    \Biggl< \biggl(
	1+e^{-\beta \bigl( U_1(\vec{x})-U_0(\vec{x}-\vec{C})-\Delta G \bigr)}
	\biggr)^{-1}
    \Biggr>_1.
    \label{eq-benn}
\end{eqnarray}
Since \dg\ is on both sides of eq (\ref{eq-benn}), it must
be solved in an iterative fashion.
Eq.\ (\ref{eq-benn}), in its un-shifted form, has been shown
to be the optimal use of bi-directional data
\cite{bennett,crooks-pre,shirts-prl,lu-jcc}.

Below we use the single-stage shifting method to calculate \dg\ using
eqs (\ref{eq-fep}), (\ref{eq-ar}) and (\ref{eq-benn})
with internal coordinates shifted by a constant $\vec{C}$,
which is chosen to maximize the overlap
between $U_0$ and $U_1$ states. For the systems studied
below, leucine dipeptide and decaglycine, we find the iterative
method of eq (\ref{eq-benn}) to be the most efficient use
of the raw data.

\subsection{\label{sec-use}Practical implementation in molecular systems} 
Determining a shifting
constant $\vec{C}$ in eq (\ref{eq-benn}) involves making
a decision about the subset of coordinates to shift. There is
generally a minimum number of coordinates needed. For example, 
for leucine dipeptide, we define the alpha and beta conformations
using two backbone torsion angles; thus these two torsions
{\it must} be shifted. In practice,
it is also possible to shift too many coordinates, leading to steric
clashes. This will be shown below for peptides. 

Once the subset of shifted internal coordinates has been
determined, the shift constant $\vec{C}$ must be calculated
for all coordinates in the subset.
In Figure \ref{fig-tors}, we chose to shift according to 
the minimum of $U_0$ and $U_1$.
This leads to our first approach: after equilibrium simulation in
each potential, find the lowest energy frame
(snapshot) for both
$U_0$ and $U_1$; then choose the constant vector $\vec{C}$ which aligns
the two lowest-energy frames,
\begin{eqnarray}
    \vec{C} = \vec{y}_0^{\rm\; min} - \vec{y}_1^{\rm\; min},
    \label{eq-ce}
\end{eqnarray}
where $\vec{y}$ represents the subset of coordinates (e.g., only
torsions) of the configuration $\vec{x}$. Strictly
speaking, $\vec{C}$ is a vector of the same dimensionality
as $\vec{x}$, with zero for every component not present in $\vec{y}$.
Another reasonable choice for a shift constant is realized by
generating a histogram for each coordinate in $\vec{y}$
for both $U_0$ and $U_1$.
The shift constant is then chosen to align the peaks of these histograms,
\begin{eqnarray}
    \vec{C} = \vec{y}_0^{\rm\; mode} - \vec{y}_1^{\rm\; mode},
    \label{eq-ch}
\end{eqnarray}
Below, we shift using both of these choices for various internal
coordinate subsets, and compare the results.

Procedurally, an estimate of \dg\ is
generated using the following steps:
\begin{enumerate}
    \item Generate an equilibrium trajectory in both states of interest,
    i.e., using $U_0$ and $U_1$.
    Determine, as discussed above, the subset of coordinates
    to shift $\vec{y}$, and the shifting
    vector $\vec{C}$ using eq (\ref{eq-ce}) or (\ref{eq-ch}).
    \item For each frame in the $U_0$ trajectory,
    shift the internal coordinates by
    $\vec{C}$ (e.g., $\phi_2 \rightarrow \phi_2+90.0$).
    Then, for each $U_0$ frame, $\vec{x}$, evaluate and record
    $U_1(\vec{x}+\vec{C})-U_0(\vec{x})$.
    \item Repeat step 2 for each frame in the $U_1$ ensemble
    with one important
    difference---the frames must be shifted in the
    opposite direction
    (e.g., $\phi_2 \rightarrow \phi_2-90.0$),
    yielding $U_0(\vec{x}-\vec{C})-U_1(\vec{x})$.
    \item Solve eq (\ref{eq-fep}), (\ref{eq-ar}) and/or (\ref{eq-benn})
    to determine a \dg\ estimate.
\end{enumerate}
Below, we utilize this process to generate multiple \dg\ estimates,
after which the mean \dg\ and standard deviation are calculated.

When shifting internal coordinates
care must be taken to shift the coordinates so that
the partition function remains unchanged. Using
the standard internal coordinates for bond length $r$, bond angle
$\theta$ and dihedrals $\omega$, a differential volume element
in configuration space is given by
\begin{eqnarray}
	d \vec{x}_1 = r_1^2 dr_1 \; \sin\theta_1 d\theta_1 \;
	    d\omega_1 
	=d\bigl(r_1^3/3\bigr) \; d(\cos\theta_1) \; d\omega_1,
    \label{eq-ham}
\end{eqnarray}
where $\vec{x}_1$ represents one possible set of internal coordinates.
Equation (\ref{eq-ham}) holds for every possible set of internal
coordinates, and thus implies that the shifting of internal coordinates
must be done according to simple rules. Bond lengths must be shifted
according to the cube of the length: $r^3 \rightarrow r^3-r_c^3$,
where $r_c$ is the bond length shifting constant
(i.e., one component of the vector $\vec{C}$) found
either by the peaks from histograms of $r^3$ and using
eq (\ref{eq-ch}), or by comparing minimum energy frames and
using eq (\ref{eq-ce}). Similarly, bond angles must be shifted 
using $\cos\theta \rightarrow \cos\theta-\cos\theta_c$,
and dihedrals are shifted via $\omega \rightarrow \omega-\omega_c$.

\section{\label{sec-results}Results for peptides}
To test the effectiveness of the
single-stage shifting method, we performed all-atom simulations of
leucine dipeptide (ACE-(leu)$_2$-NME)
and decaglycine (ACE-(gly)$_{10}$-NME).
Both peptides were simulated using the TINKER Version 4.2
molecular dynamics package \cite{ponder}.
The peptides were solvated implicitly
using the generalized Born surface area (GBSA)
approach \cite{still} and Langevin dynamics were utilized with
the friction coefficient set to that of water (91.0 psec$^{-1}$).
A time step of 1.0 fsec was chosen for all simulations.
Leucine dipeptide was maintained at
500.0 K (to enable independent verification of our result)
and utilized the CHARMM27 forcefield parameter set \cite{charmm}.
Decaglycine was maintained at 300.0 K and used the AMBER96 forcefield
parameter set \cite{amber} for comparison with Ref.\ \citenum{meirovitch-deca}.

\subsection{Leucine dipeptide}
Leucine dipeptide (ACE-(leu)$_2$-NME)
was chosen as a test system because
(i) it possesses some of the complexity of a large
molecule---four backbone torsions and eight side-chain
dihedrals; and
(ii) it is small enough to allow for very long simulation times.
Long simulation times are necessary
for an unbiased, independent determination of the
conformational population (hence \dg), providing a
strict test of the single-stage shifting method.

For leucine dipeptide we calculated the free
energy difference \dg\ and entropy difference \ds\ for
the alpha $\rightarrow$ beta conformational change.
We defined the alpha and beta conformations using two internal
backbone torsions, namely,
for alpha: $-145<\phi_2<-25 \;{\rm and}\; -125<\psi_1<-5$;
and for beta: $-160<\phi_2<-40 \;{\rm and}\; 70<\psi_1<-170$.
A temperature of 500 K was chosen to enable repeated crossing of
the free energy barrier between the alpha and beta conformations.
At 500 K, leucine dipeptide switches between alpha and beta conformations
at a rate of around 2.5 transitions per nsec with GBSA solvation.
Note that this high temperature is required only to obtain an unbiased
estimate of \dg---our single-stage shifting approach works
equally well at lower temperatures.

To obtain an independent and unbiased value of
\dg, four 1.0 $\mu$sec simulations
were performed yielding around 2500 transition events per trajectory.
The four trajectories were then used to calculate \dg\ via the definition
\begin{eqnarray}
    \Delta G \equiv -\frac{1}{\beta}
	\ln\Biggl( \frac{N_{\rm beta}}{N_{\rm alpha}} \Biggr),
    \label{eq-count}
\end{eqnarray}
where $N_{\rm alpha}$ and $N_{\rm beta}$ are, respectively, the number
of dynamics steps the system was in the alpha and beta conformations.
The unbiased value of \dg\ for alpha $\rightarrow$ beta was found to be
\dg\ $=0.95 \pm 0.05$ kcal/mole, where the value of \dg\ given
is the average of the four 1.0 $\mu$sec estimates with
uncertainty given by the standard deviation.

The unbiased entropy difference was calculated using
eq (\ref{eq-count}) via
$T\Delta S = \Delta U_{\rm avg}-\Delta G$ where
$\Delta U_{\rm avg}$ is the average of
$U_{\rm beta}-U_{\rm alpha}$.
We found that $T$\ds\ was zero within uncertainty.
This is consistent with our observation
that the fluctuating degrees of freedom for leucine dipeptide are
mainly the side-chain torsions, and thus
do not change dramatically between
the alpha and beta conformations. Given the overall \dg\ value,
the alpha conformation is favored to to intra-molecular attractions.

With an unbiased value of \dg\ from eq (\ref{eq-count}), it is possible to
test various implementations of the single-stage shifting method.
To this end, we simulated leucine dipeptide in the alpha and beta
conformations, generating four 1.0 nsec trajectories for each conformation.
Each of the trajectories were obtained by constraining the
backbone torsions to stay within the defined ranges for alpha and beta,
given above. The constraining force was zero if the torsion
angles were within the allowed range, and harmonic otherwise.
Frames were saved every 0.1 psec yielding 10,000 frames per trajectory.

The four trajectories provide sixteen independent
single-stage shifting estimates of \dg\ and
\ds\ using all possible pairings.
The results for \dg\ are summarized in Table \ref{tab-dileu} 
where we used both the lowest energy frame and histogram
peak shifting approaches (Section \ref{sec-use}) to estimate \dg\
for leucine dipeptide. The value of \ds\ was found to be zero within
uncertainty, consistent with the value found using eq (\ref{eq-count}).
For each shifting approach,
we tested four sets of shifting coordinates: backbone torsions only,
all torsions, all
torsions and bond angles, or all internal coordinates (torsions, angles
and bond lengths). In addition, we also show results using both
the iterative method of eq (\ref{eq-benn}) and the acceptance-ratio
method of eq (\ref{eq-ar}).
The values for \dg\ are averages with 
standard deviations shown in parentheses.
Our single-stage shifting results in Table \ref{tab-dileu} agree
well with the unbiased estimate given above.
The table also show that shifting torsions
results in a small uncertainty,
while including bond angles and lengths makes the \dg\
estimate less certain.

In all of our simulations, we have found that shifting torsions
(either backbone only, or all torsions), and using
the iterative method of eq (\ref{eq-benn}),
has consistently provided
accurate results. Also, as demonstrated in Table \ref{tab-dileu},
lower uncertainty is obtained by shifting according
to histogram peaks and using eq (\ref{eq-ch})
rather than shifting by the lowest energy frames.

\begin{figure}
    \begin{center}
	\includegraphics[scale=0.15,clip]{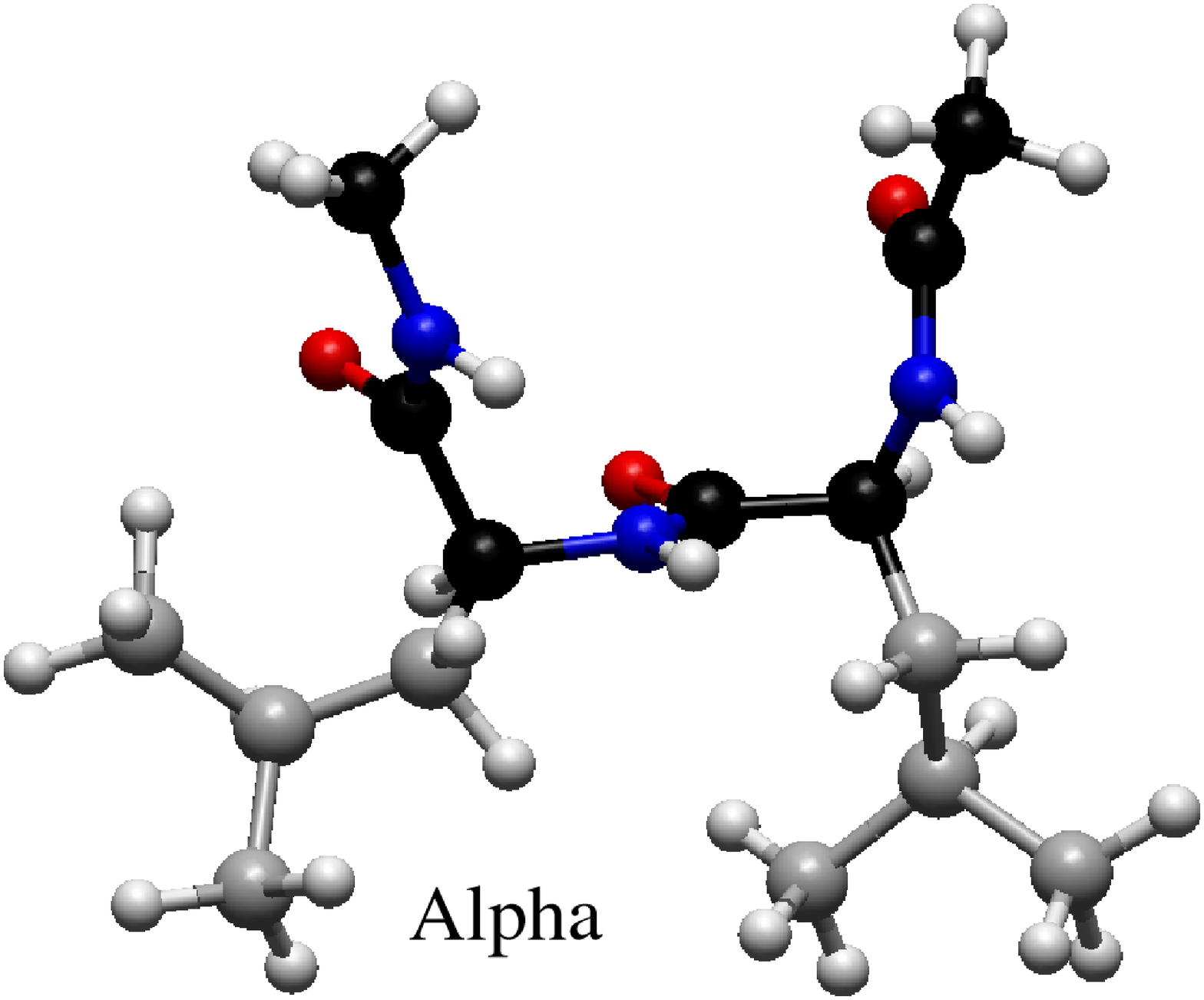}
	\includegraphics[scale=0.15,clip]{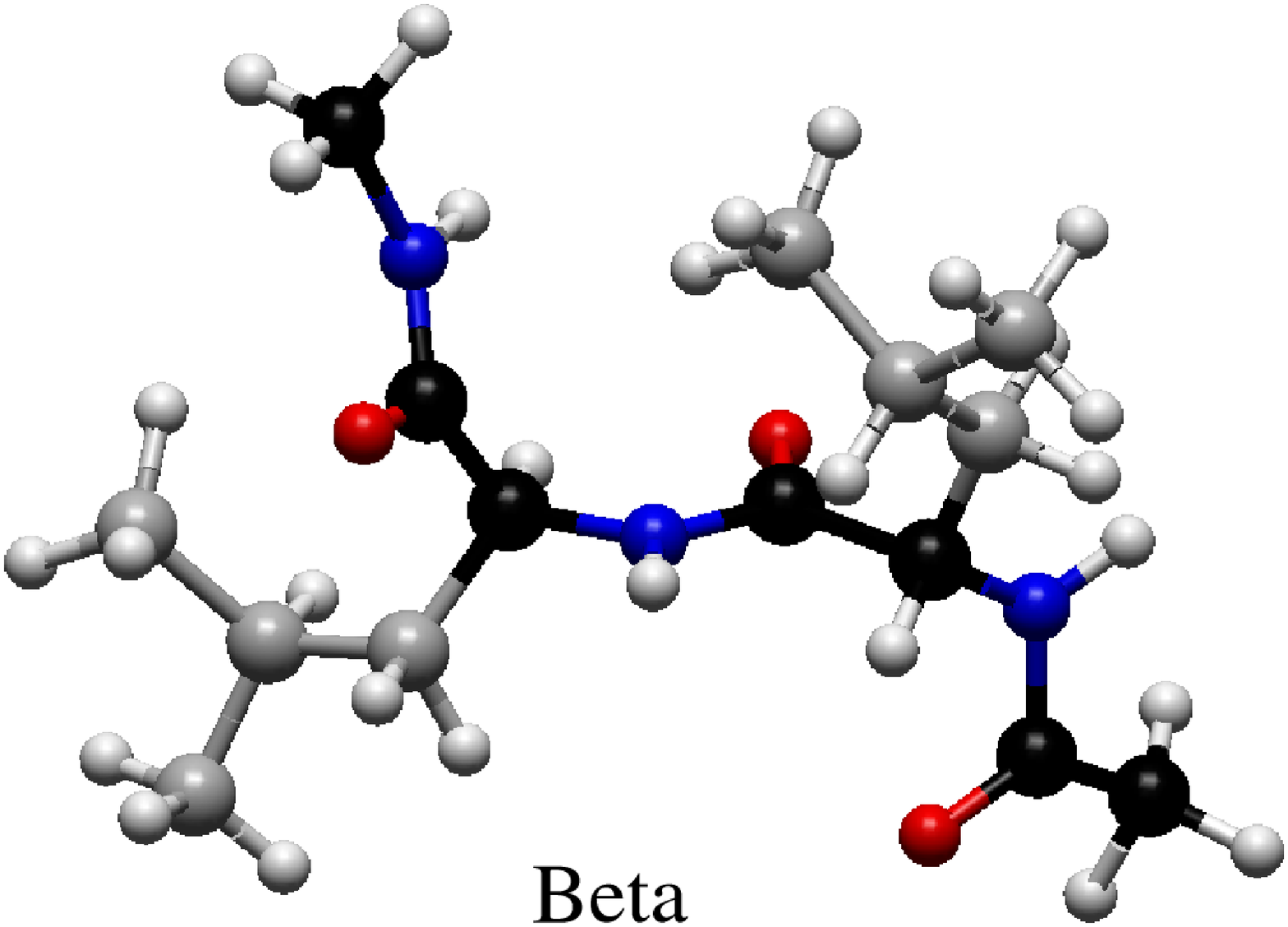}\\
	\includegraphics[scale=0.45,clip]{dileu-compare.eps}
    \end{center}
    \caption{\label{fig-dileu-compare}
	Comparison between several shifting methods
	for the conformational change in free energy 
	for alpha $\rightarrow$ beta of leucine dipeptide in GBSA solvent.
	The alpha (top left, ``cis-like'') and beta
	(top right, ``trans-like'') conformations
	are also shown, with the backbone carbons colored black.
	Two trajectories (one in each conformation) were analyzed
	using several protocols of the single-stage shift technique.
	The solid horizontal black line represents the 4.0 $\mu$sec value
	of \dg\ obtained by using eq (\ref{eq-count}).
	The curves are obtained using the single-stage shifting method
	where backbone torsions were shifted according to the peak
	of the histogram (i.e., lowest uncertainty \dg\ estimate
	in Table \ref{tab-dileu}).
	The red curve was generated from
	Bennett's acceptance-ratio method,
	the green dashed curve is free energy perturbation in the forward
	direction (i.e., $\Delta G_{\rm alpha \rightarrow beta}$),
	and the green
	solid curve is free energy perturbation in the reverse direction
	(i.e., $-\Delta G_{\rm beta \rightarrow alpha}$). The solid blue curve
	is Bennett's iterative method.
	The figure clearly suggests the superior convergence properties
	of the iterative method for our single-stage shifting data.
    }
\end{figure}

We also stress the importance of using bi-directional
data to determine \dg\ for conformational equilibria. 
To this end, in Figure \ref{fig-dileu-compare},
we employed the single-stage shifting method using
three data analysis protocols:
single-stage free energy perturbation in both directions
(eq (\ref{eq-fep})) \cite{zwanzig},
Bennett's acceptance-ratio method (eq (\ref{eq-ar}))
and Bennett's iterative method (eq (\ref{eq-benn})) \cite{bennett}.
The superior convergence properties of the iterative method
can be seen in Figure \ref{fig-dileu-compare}. The
solid horizontal black line represents the independent,
4.0 $\mu$sec \dg\ value obtained by
using eq (\ref{eq-count}).
The data are \dg\ estimates using the single-stage shifting method,
where backbone torsions were shifted according to the histogram peaks.
The data for the figure was generated using a single trajectory in
each of the alpha and beta conformations.
The red curve was generated from Bennett's acceptance-ratio method,
the green dashed curve is free energy perturbation in the forward
direction (i.e., $\Delta G_{\rm alpha \rightarrow beta}$), and the green
solid curve is free energy perturbation in the reverse direction
(i.e., $-\Delta G_{\rm beta \rightarrow alpha}$). Finally the solid blue curve
is Bennett's iterative method.
It is clear from the figure that using bi-directional data is very
important to the success of our single-stage shifting method. Further,
Bennett's iterative method is shown to converge more quickly than
the other methods.

\begin{figure}
    \begin{center}
	\includegraphics[scale=0.5,clip]{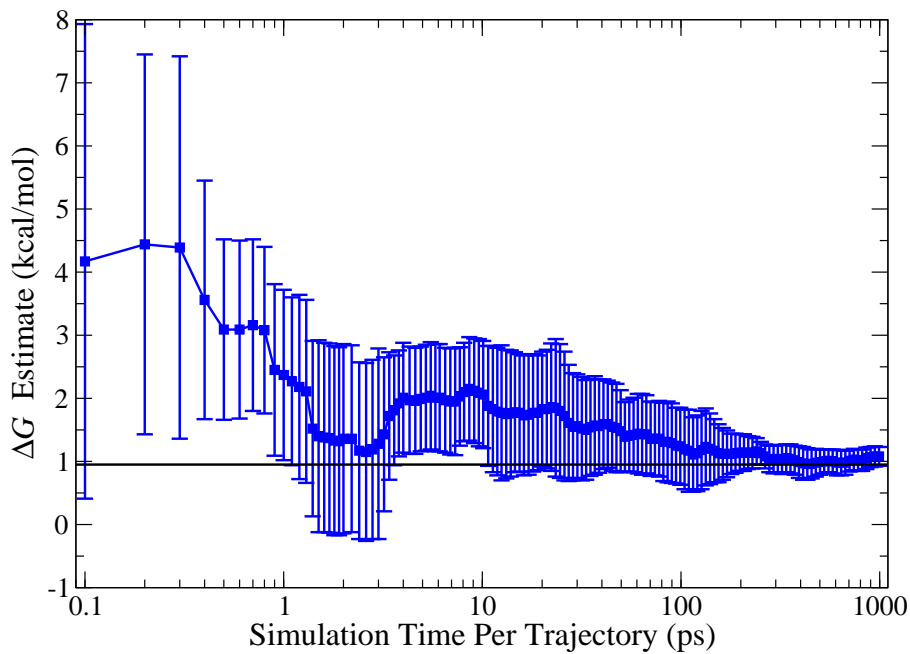}
    \end{center}
    \caption{\label{fig-dileu-df}
	Conformational free energy difference \dg\
	for alpha $\rightarrow$ beta conformations
	of leucine dipeptide in GBSA solvent,
	shown as a function of the simulation time.
	The horizontal black line shows the known \dg\ obtained using
	4.0 $\mu$sec of simulation and eq (\ref{eq-count}).
	The data were obtained using the single-stage shifting method
	where backbone torsions were shifted according to the
	peak of the histogram, and the iterative
	method of eq (\ref{eq-benn}) was used to analyze the data.
	The average \dg\ estimates (blue squares)
	and standard deviations (errorbars) were calculated using sixteen
	independent estimates of \dg.
	Using the single-stage shifting method, with only 30 psec of
	simulation (15 psec in alpha and 15 psec in beta), the average
	estimate of \dg\ is within 1.0 kcal/mol of the known value
	with a standard deviation of less than 1.0 kcal/mol.
	This may be compared with unconstrained simulations,
	in which transitions between
	alpha and beta conformations occurred, on average, every 400 psec.
    }
\end{figure}

Figure \ref{fig-dileu-df} demonstrates the
efficiency of the single-stage shifting method compared to
long simulation and use of eq (\ref{eq-count}).
The horizontal black line is the unbiased \dg\ from
long (4.0 $\mu$sec) simulation. The curve shows the average
(blue squares) and standard deviation (errorbars)
of the single-stage shifting method where backbone torsions
were shifted by the histogram peaks, and the iterative
method of eq (\ref{eq-benn}) was used to analyze the data.
In our unconstrained simulations, leucine
dipeptide switched between the alpha and beta conformations,
on average, once every 400 psec. Using our single-stage
shifting method, with only 30 psec of simulation (15 psec in
alpha and 15 psec in alpha), a reasonably
accurate and precise value of \dg\ can
be obtained.

\subsection{Decaglycine}
We also applied the single-stage shifting method to decaglycine 
(ACE-(gly)$_{10}$-NME), predicting
the conformational \dg\ and entropy difference \ds\
for alpha $\rightarrow$ extended conformations.
We again defined the alpha and extended conformations by the
internal backbone torsions
(i.e., excluding $\phi_1$ and $\psi_{10}$), namely,
for alpha: $-115<\phi<5 \;{\rm and}\; -115<\psi<5$;
and for extended: $120<\phi<-120 \;{\rm and}\; 120<\psi<-120$.
Previous \dg\ and \ds\ calculations of decaglycine in vacuum
were performed by Karplus and Kushick \cite{karplus-deca},
and quite recently by Cheluvaraja and Meirovitch \cite{meirovitch-deca}.
Apparently, decaglycine's conformational \dg\ and \ds\ have
not previously been computed in implicit solvent.

To calculate \dg\ and \ds, four trajectories
in each conformation were
generated using the simulation parameters defined previously.
Thus, sixteen independent \dg\  and \ds\ estimates can be calculated.
Each trajectory was 1.0 nsec in length with a frame saved every
0.1 psec yielding 10,000 frames per trajectory---although this may be
quite sub-optimal; see Section \ref{sec-opt}.
As with leucine dipeptide,
the sixteen \dg\ and \ds\ estimates were generated using
various shifting approaches.

Table \ref{tab-decagly} shows the results of our calculation
of \dg\ and \ds\ for the alpha $\rightarrow$ extended conformational
change using various shifting approaches.
The entropy change was estimated via
$T\Delta S = \Delta U_{\rm avg}-\Delta G$ where
$\Delta U_{\rm avg}$ is the average of $U_{\rm ext}-U_{\rm alpha}$.
Acceptance ratio estimates had a much larger uncertainty
then the iterative method estimates and thus are not included
in the table. The results clearly demonstrate that shifting by
histogram peaks provides a higher level of precision than shifting
by the lowest energy frame.
Also, as with leucine dipeptide, lower uncertainty is obtained
when shifting torsions only (i.e., not bond angles and lengths).

The results in Table \ref{tab-decagly} suggest that the
``compensating''
role of the entropy is vital for an accurate \dg\ calculation
in GBSA solvent. 
Our studies of decaglycine in vacuum (data not shown), as well
those of other groups \cite{meirovitch-deca,karplus-deca} show
that the alpha conformation is more stable than extended,
due mainly to energetics.
However, our results suggest that, with the addition of (implicit)
solvent, the energy difference between the two conformations becomes
small enough that the entropy term dominates \dg---to the degree
that the extended
conformation is more stable than alpha.

\begin{figure}
    \begin{center}
	\includegraphics[scale=0.1,clip]{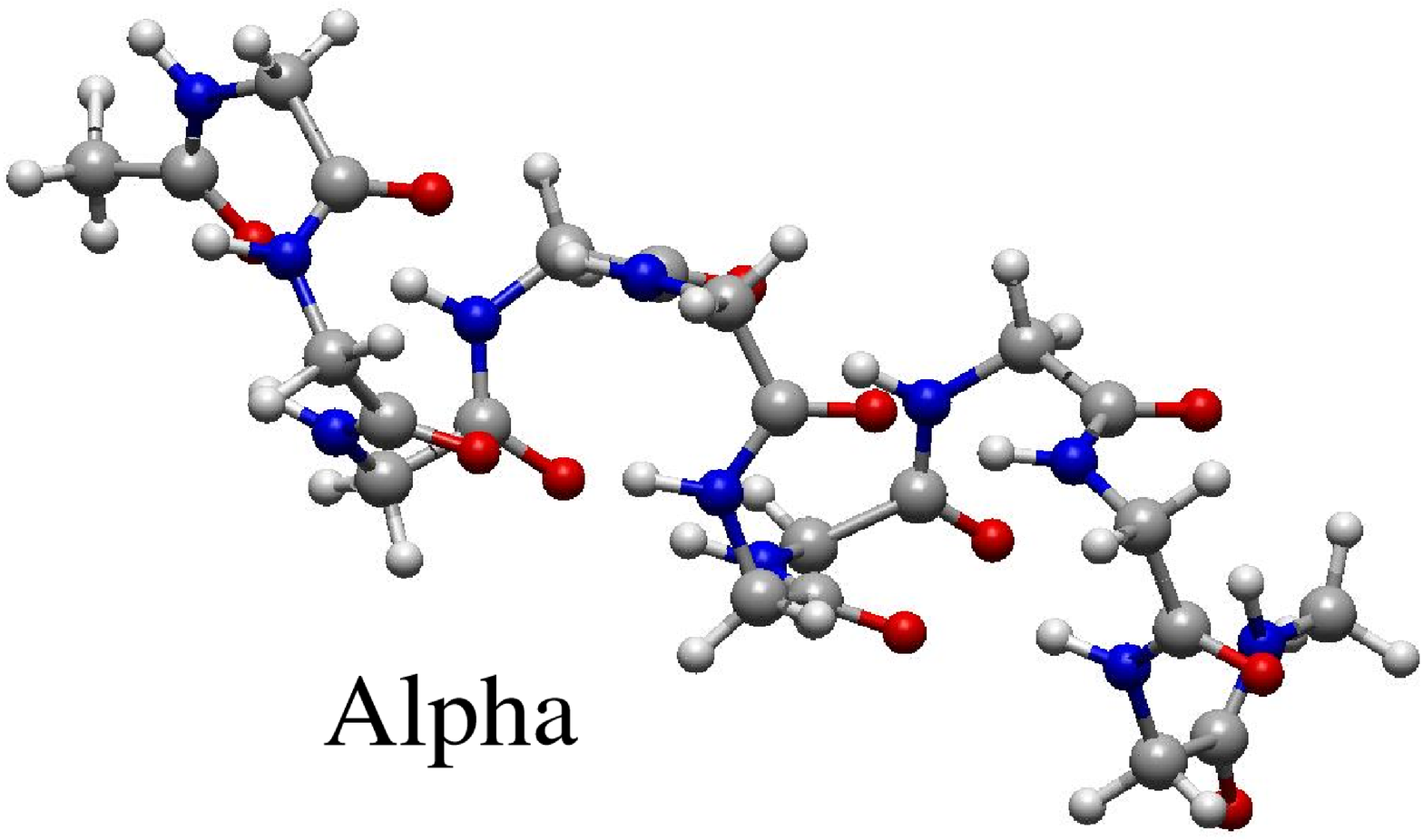}
	\includegraphics[scale=0.12,clip]{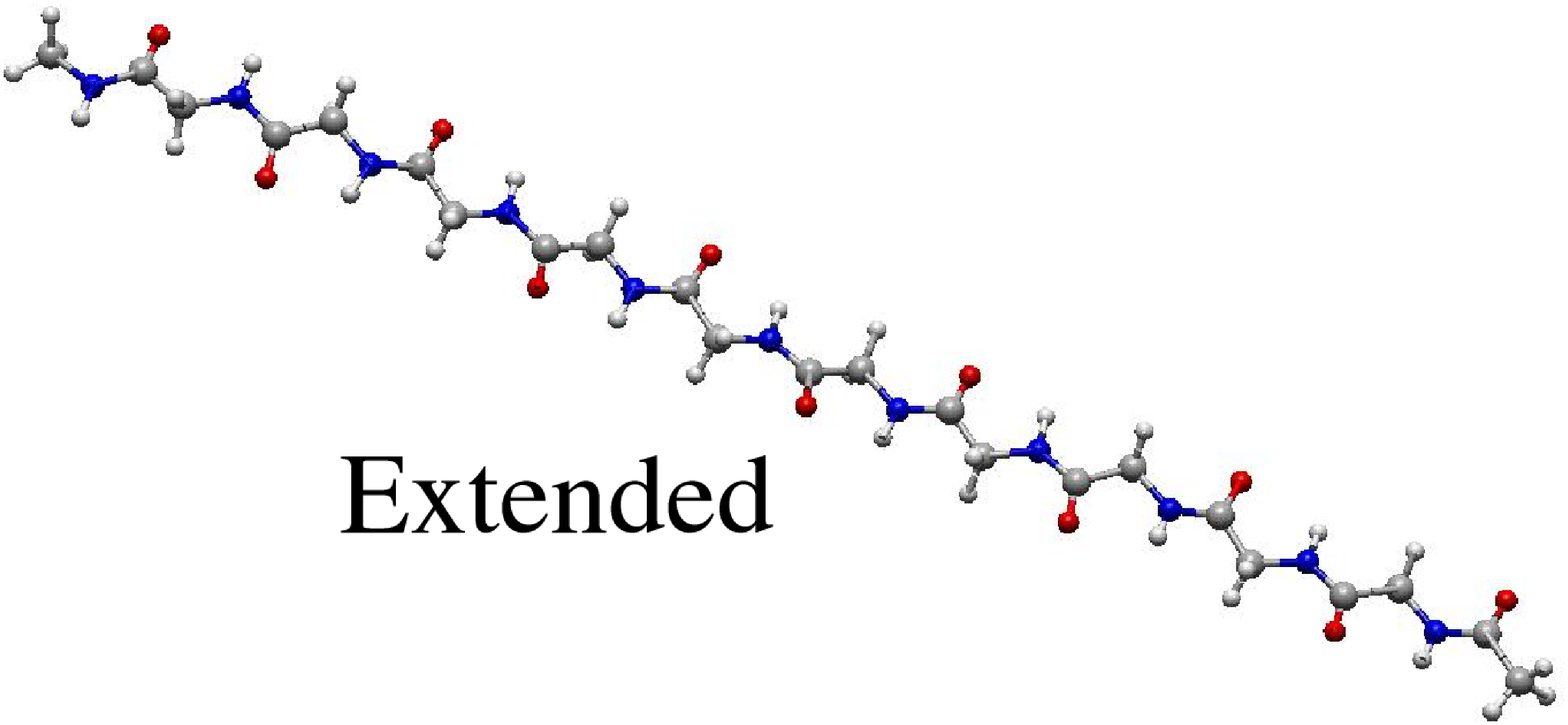}\\
	\includegraphics[scale=0.33,clip]{decagly-df.eps}
	\includegraphics[scale=0.33,clip]{decagly-ds.eps}
    \end{center}
    \caption{\label{fig-decagly}
	Conformational free energy difference \dg\ and
	entropy difference $T$\ds\ for the 
	alpha $\rightarrow$ extended conformational change in
	decaglycine with GBSA solvent.
	The alpha (top left) and beta (top right) conformations
	are also shown.
	The data were obtained from eight independent simulations---four
	constrained to be in alpha and four in extended.
	From the eight simulations, sixteen independent
	estimates of \dg\ were obtained, and
	the averages (blue squares) and standard deviations
	(error bars) were calculated as a function of 
	simulation time.
	Each \dg\ estimate was generated using the end point
	shifting method by shifting backbone torsions via
	histogram peaks, and using the iterative method
	of eq (\ref{eq-benn}).
    }
\end{figure}

Figure \ref{fig-decagly} shows \dg\ and $T$\ds\
as a function simulation time.
The data points are the average of the sixteen independent
estimates with standard deviation given by the error bars for both
\dg\ (blue squares) and $T$\ds\ (green circles).
These estimates were obtained using the iterative method,
and shifting the backbone torsions by the histogram peak.
The figure demonstrates the apparent convergence of the
\dg\ and $T$\ds\ estimates. Note that it would
be impractical to obtain an independent estimate for decaglycine
(as we did with leucine dipeptide) because of the required
simulation times.

In the current implementation, a total of 8 nsec of simulation was
required to obtain a reasonably accurate and precise estimate of \dg.
To our knowledge, no multi-stage calculation has ever been
attempted on this system, undoubtedly due to the prohibitive computational
expense. Nevertheless, we believe additional optimization of our current
shifting protocol will be possible, as we now discuss.

Finally, we note that, as in any \dg\ computation, our results reflect
the definitions chosen for the alpha and extended states.

\section{\label{sec-opt}Further optimization of the single-stage
         shifting approach}
While the present results indicate that peptide conformational equilibria
can be determined by sub-nsec simulations, several promising avenues
for optimization have not been explored. We briefly sketch several
possible approaches for improving efficiency,
including combining the shifting approach with
traditional staged calculations.

It is useful to consider the upper limit for the computational
cost of the single-stage shifting
procedure. The maximum is essentially twice that of
equilibrium simulation,
provided the method is hard-wired into the molecular dynamics
program, due to the extra energy call
that must be made for each shifted frame.
In the current study, the shifting procedure was scripted
external to the simulation program (TINKER), and
thus the cost for trajectory analysis was high---limiting the
number of frames per trajectory to 10,000.
We found that, for this fixed number of frames per trajectory, the
simulation time between frames had very little affect on the
\dg\ estimate. Thus, we feel that substantial increases in efficiency
could be realized by utilizing {\it every} frame in the
\dg\ calculation---i.e., every time step.
(It is worthwhile to recall that interactions change
enough over a single time step to require
re-calculation of forces.)
Ultimately, then, reliable and accurate \dg\
calculations should take
no more time than required to sample the equilibrium
ensemble in a given state
(the minimum time for any \dg\ method).

Additionally, the shifting procedures explored in this report
ignore correlation between coordinates, such as those known
from Ramachandran plots,
where backbone torsions $\phi$ and $\psi$ do not vary independently.
Ramachandran plots, moreover, average over many residues, which
individually are likely more correlated. To motivate
more general shifts, consider one
state where a certain pair of $\phi,\psi$ angles
inhabit a predominantly
vertical region of Ramachandran space, while the other state
populates a region with a very different
orientation. In such a case, simple shifts
alone (e.g., those in eqs (\ref{eq-ce}) and (\ref{eq-ch}))
will not
maximize overlap to the extent that a combined shift and
(partition-function-preserving) rotation would.
Further, if one oblong well is very narrow and the other
well is very broad, then coordinate scaling (contraction/expansion)
should also be performed
(see also Refs.\ \citenum{jarzynski-targeted,reinhardt-scaling}).
In general the coordinates can be linearly scaled,
rotated and/or translated using a constant matrix $\bf A$
(i.e., $\vec{x} \rightarrow {\bf A} \vec{x} + \vec{C}$),
and eqs (\ref{eq-fep}),
(\ref{eq-ar}) and (\ref{eq-benn}) must be generalized
to account for the matrix $\bf A$.
More complex, nonlinear transformations are also possible,
but may not be practical.

In larger systems than those considered here (e.g., whole proteins),
the gain in overlap due to the internal coordinate shift may prove
insufficient to permit reliable single-stage computation of \dg\
values. In such cases, it may be advantageous to
combine the single-stage shifting approach with multi-stage
methodology. To do so, a path connecting
the two states of interest can be defined
(e.g., Refs.\ \citenum{brooks-diala,roux,brooks-ldyn}),
and independent trajectories can be generated
at intermediate stages along the path.
Then, between each successive intermediate stage,
the incremental free energy difference ($\delta G$) is
estimated using the shifting protocol outlined in Section
\ref{sec-use}. The $\delta G$ estimates can then
be summed to obtain the full \dg\ for the complete path.

\section{\label{sec-ext}Extension to ``alchemical''
    calculations and explicitly solvated systems}
It is possible to extend the formalism of the single-stage shifting
method beyond conformational \dg\ calculation for implicitly
solvated molecules.
In this section we outline the potential for the single-stage
shifting method to be used for ``alchemical'' mutations---which
are the basis for relative binding affinities
and solubilities \cite{tembe}---and 
on explicitly solvated systems.

For alchemical mutation, two distinct potential energy functions $U_0$
and $U_1$---one for each molecule---are used in eqs
(\ref{eq-fep}), (\ref{eq-ar}) and (\ref{eq-benn}).
While the mathematical formalism is unchanged from
conformational \dg\ calculations,
the difficulty in alchemical mutations lies in
determining the shifting vector
$\vec{C}$, since the number of degrees of freedom for $U_0$
and $U_1$ are different, in general. (For conformational
\dg\ calculations, such as those above, the number of degrees
of freedom for $U_0$ and $U_1$ are always the same.)
This difficulty can be overcome by introducing ``dummy''
coordinates as in Ref.\ \citenum{roux}.
Although dummy coordinates will change the absolute
free energy values, use of a thermodynamic cycle guarantees
that the free energy difference \dg\ will
remain unchanged.
Thus, accurate {\it relative} binding affinities and solubilities can
be obtained \cite{roux}.

If explicit solvation is used for conformational or alchemical \dg\
calculation, then the single-stage shifting method must be generalized
to include non-standard
{\it inter}molecular ``external'' coordinates.
This can, in principle, be accomplished by
introducing auxiliary vectors
to describe the location and orientation (e.g., Euler angles)
of each solvent or solute molecule. The necessary shifting vector
$\vec{C}$ will now include the full set of intra- and inter-molecular
coordinates.

\section{Conclusion}
In a study of peptide equilibria, we have
demonstrated a simple method for substantially
overcoming the overlap problem in calculating free
energy differences (\dg) and entropy difference
(\ds) between conformational states.
The new single-stage shifting method utilizes
a shift in internal coordinates to improve the overlap
between configurations, motivated by Voter's study in
Ref.\ \citenum{voter}.
The approach requires only simulation in the two states of interest
without the need for ``staged'' intermediate calculations.
Bennett's iterative method \cite{bennett}
is used to efficiently calculate a \dg\ value from the raw
data.

We tested the single-stage shifting approach on two
peptides, obtaining
excellent results with sub-nsec simulation times.
First, for leucine dipeptide in implicit
solvent, we accurately calculated the conformational \dg\ for
alpha $\rightarrow$ beta conformations---judging by
nearly perfect agreement with a 4.0 $\mu$sec simulation.
The entropy difference \ds\ was found to be nearly
zero, also consistent with long simulation.
The single-stage shifting method was then
used to predict the conformational 
\dg\ and \ds\ for
alpha $\rightarrow$ extended conformations of decaglycine
in implicit solvent, apparently for the first time.
We find that, with implicit solvent, the extended conformation
of decaglycine is favored over alpha, due mainly to the entropy
gain in the extended state.
By contrast, in vacuum, the alpha conformation is
preferred due mainly to the strongly favorable intra-molecular
interactions.
It must be borne in mind that our quantitative results
necessarily depend on our state definitions.

While the present report describes a single type of application
of the single-stage shifting approach (to conformational equilibria),
we believe the
idea will find quite broad applications---in part due to
the substantial potential for further optimization.
We have therefore discussed optimization of the method,
and application to ``alchemical'' mutations (for
relative binding affinities), as well as the use of explicit solvent.
The single-stage shifting approach may also be combined
with multi-stage simulation, allowing further optimization.
We are currently exploring these ideas.

\section*{Acknowledgments}
We would like to thank Carlos Camacho,
Ronald White, Srinath Cheluvaraja
and Edward Lyman for many fruitful discussions.
Funding for this research was provided by
the Dept.\ of Environmental and Occupational
Health and the Center for Computational Biology
and Bioinformatics at the University of Pittsburgh, and
the National Institutes of Health (Grant T32 ES007318).

\bibliography{/home/marty/res/other/tex/bib/my}

\clearpage
\begin{table}
    \begin{center}
    \begin{tabular}{l|l|c|c}
	\hline \hline
	Shifting  &                     & Iterative       & Accept.\ ratio \\
	approach  & Shifted coordinates & \dg\ (kcal/mol) & \dg\ (kcal/mol) \\
	\hline
	Peak of   & Backbone torsions only       & 1.08 (0.15) & 1.02 (0.13) \\
	histogram & All torsions                 & 0.56 (0.87) & 0.55 (0.86) \\
	(eq (\ref{eq-ch})) & All torsions and bond angles & 0.76 (2.64)
							    & 1.01 (3.25) \\
	        & All internal coordinates     & 0.78 (2.63) & 1.09 (3.26) \\
	\hline
	Lowest    & Backbone torsions only       & 1.29 (0.35) & 1.30 (0.39) \\
	energy    & All torsions                 & 0.99 (0.66) & 0.98 (0.66) \\
	frames    & All torsions and bond angles & 3.36 (4.75) & 6.48 (9.45) \\
	(eq (\ref{eq-ce})) & All internal coordinates     & 4.50 (7.16)
							    & 8.80 (14.32) \\
	\hline \hline
    \end{tabular}
    \end{center}
    \caption{\label{tab-dileu}
	Conformational free energy differences \dg\ for
	alpha $\rightarrow$ beta conformations for leucine
	dipeptide in GBSA solvent using several
	variations of the single-stage shifting approach.
	Each shifting \dg\ result is an average of sixteen
	independent 2.0 nsec estimates of \dg\ with
	standard deviation shown in
	parentheses. The iterative results use eq (\ref{eq-benn})
	and the acceptance-ratio results use eq (\ref{eq-ar}).
	For comparison, the value of \dg\
	obtained from long simulation (4.0 $\mu$sec) is 0.95 (0.05) kcal/mole.
	The entropy difference \ds, found by single-stage shifting,
	was zero within uncertainty, consistent with the value
	obtained by long simulation.
	This table also demonstrates that shifting too many coordinates
	(i.e., bond angles and lengths) leads to steric clashes, and
	thus a larger uncertainty for \dg.
    }
\end{table}

\clearpage
\begin{table}
    \begin{center}
    \begin{tabular}{l|l|c|c}
	\hline \hline
	Shifting &                     & Iterative       & Iterative \\
	approach & Shifted coordinates & \dg\ (kcal/mol) & $T$\ds\ (kcal/mol) \\
	\hline
	Peak of   & Backbone torsions only       & -12.39 (0.47) 
							& 9.75 (0.66) \\
	histogram & All torsions                 & -12.59 (0.82)
							& 9.95 (0.93) \\
	(eq (\ref{eq-ch})) & All torsions and bond angles & -12.77 (2.94)
							& 10.12 (2.97) \\
	          & All internal coordinates     & -12.75 (3.11)
							& 10.10 (3.13) \\
	\hline
	Lowest    & Backbone torsions only       & -11.97 (1.59)
							& 9.34 (1.71) \\
	energy    & All torsions                 & -12.12 (2.57)
							& 9.48 (2.76) \\
	frames    & All torsions and bond angles & -18.53 (11.14)
							& 15.89 (10.93) \\
 	(eq (\ref{eq-ce})) & All internal coordinates     & -24.79 (14.03)
							& 22.15 (13.86) \\
	\hline \hline
    \end{tabular}
    \end{center}
    \caption{\label{tab-decagly}
	Conformational free energy differences \dg\ and entropy
	differences $T$\ds\ for alpha $\rightarrow$ extended
	conformations of decaglycine
	in GBSA solvent using several shifting approaches.
	Each shifting result is an average of sixteen 
	independent 2.0 nsec estimates with standard deviation shown in
	parentheses, using the iterative method of
	eq (\ref{eq-benn}). Acceptance ratio results (data not shown)
	had a much larger uncertainty than the iterative method.
	This data clearly shows that lower uncertainty
	is obtained by shifting via histogram peaks rather than
	by the lowest energy frame. Also, shifting non-torsional
	coordinates (i.e., bond angles and lengths)
	dramatically increases uncertainty.
    }

\end{table}

\clearpage

\clearpage

\clearpage

\clearpage

\end{document}